# Conductance spectrum and superconducting gap structures observed in c-axis FeSe$_{0.3}$Te$_{0.7}$/Au junctions

Y. T. Shen[1], Y. S. Li[2], K. C. Lin[1,2], M. K. Wu[2,3], C. C. Chi[1]

[1]*Department of Physics, National Tsing Hua University, Hsinchu 300, Taiwan*
[2]*Institute of Physics, Academia Sinica, Nankang, Taipei 115, Taiwan*
[3]*Department of Physics, National Dong Hwa University, Hualien 974, Taiwan*

**ABSTRACT**

The electric transport properties of superconducting FeSe$_{0.3}$Te$_{0.7}$/Au c-axis (S/N) junctions, fabricated by using Pulsed Laser Deposition technique, have been investigated in the temperature range of 2 K to the superconducting transition temperature T$_C$ ~ 13.5 K, and in the presence of applied magnetic fields from 0 to 9 T. A large zero bias conductance peak has always been observed in the conductance spectra of several batches of junctions. In addition, we have also found several gap-like features. Using the extended BTK theory with the currently favored nodeless $S_\pm$ − wave symmetry, our conductance spectrum can be reproduced quite well in the low bias range with $\Delta_1 = 4\,\text{meV}$, and $\Delta_2 = 6\,\text{meV}$ at 2 K. However the experimental conductance spectrum is substantially below the calculated one in the high bias range. Furthermore, there is a conductance minimum at about 20 meV, which may be the reason for the discrepancy.

# I. INTRODUCTION

The discovery of a new superconductor, fluorine-doped LaFeAsO (the so-called 1111 system), with a $T_C$ of 26 K by Hideo Hosono et al. [1] in 2008 was a great surprise and triggered world-wide efforts to understand the nature of this new class of superconducting compounds. Soon after the discovery, various iron-based superconductors with similar layered structures, such as MFeAs (the 111 system, where M stands for alkaline metal) and $AeFe_2As_2$ (the 122 system, where Ae stands for alkaline-earth metal) have been reported [2- 5]. These and some other variations containing iron and arsenic are generally called iron arsenide superconductors. Shortly after the discovery of iron arsenide superconductors, M.K. Wu and his collaborators have discovered the iron chalcogenide superconductors, among which FeSe (the 11 system) has the simplest structure, because it contains only the FeSe sheet without any inter-layers [6]. But the superconducting transition temperature of FeSe is relatively low at about 8 K. Subsequent studies of this 11 system have revealed that its $T_C$ is extremely sensitive to applied pressure and can rise to a value of 37 K at 8.9 GPa [7- 10]. Moreover, by partial substituting Se with Te, $T_C$ can be enhanced up to a maximum of 14 K [11, 12].

Despite the intense investigation on these iron-based superconductors, the pairing symmetry in these compounds remains unsettled. Theories of several kinds of gap symmetries such as p-wave [13], d-wave [14], and $S_\pm$ − wave [15- 19] have been proposed, and the $S_\pm$ − wave pair symmetry is the currently favored candidate. The $S_\pm$ − wave depicts that there are two s-wave order parameters with a opposite sign between them, one for the hole Fermi surfaces centered at Γ point, and one for the electron Fermi surfaces centered around M point in the Brillouin zone. For the iron arsenide superconductors, the S±-wave scenario is supported by the phase sensitive superconducting quantum interference device (SQUID) measurement [20]. However, for the iron chalcogenide series, whether the $S_\pm$ − wave is the correct symmetry remains controversial [21, 22].

In this paper, we report the transport measurements in the c-axis S/N junction. We have observed a prominent zero bias conductance peak (ZBCP) with a small peak splitting at low temperatures and some gap-like features below $T_C$. The conductance spectra seem consistent with the $S_\pm$ model. In addition, we have also found an interesting $T_C$ enhancing effect by post-annealing the pulse-laser-deposited FeSeTe thin films in oxygen. Similar $T_C$ enhancement for single crystal FeSeTe has been reported by Hefei Hu et al [23].

## II. FABRICATION OF FESETE THIN FILMS

We used the pulse laser deposition (PLD) technique to grow the FeSeTe thin films on (100) Magnesium Oxide (MgO) substrate in a vacuum chamber with a base pressure better than $5 \times 10^{-5}$ mbar. A pulsed KrF ($\lambda = 248$ nm) excimer laser (Lambda Physik LPX) was used to bombard the target with a nominal composition of $FeSe_{0.3}Te_{0.7}$. We have varied the deposition conditions to optimize the $T_C$ of FeSeTe thin films, and found that the optimal deposition temperature, laser pulse energy density, and the target-to-substrate distance are $350°C$, 1 J/cm², and 48 mm respectively.

An interesting $T_C$ enhancing effect was accidentally discovered when we repeat the resistance-versus-temperature measurements of the same sample after having it exposed to the air at room temperature for different days as shown in Fig. 1(a). The R(T) data was obtained by using the standard four-probe method. The black squares, the red solid circles, the blue open circles, and the cyan crosses in Fig. 1(a) correspond to the R(T) data for the as-grown sample, one day, three days, and five days in air respectively. The fresh as-grown sample exhibits an onset superconducting transition temperature of 11.9 K with a transition width around 2.2 K. A substantial increase in $T_C$ with reduction in transition width was observed for the sample exposed in air for just one day. The onset transition temperature reached a maximum of 14.2 K after exposing in air for 3 days, and no significant increase in $T_C$ was found for longer air exposure.

We speculate that the enhancement of superconductivity is due to the oxygen in the air. Therefore, we perform an in-situ oxygen annealing process right after depositing FeSeTe film. The as-grown sample was annealed at $70°C$ in 1 atm $O_2$ for half day. The transition temperature and transition width for this $O_2$ annealed thin film became 14.5 K and 1.6 K respectively, as shown in Fig. 1 (b). This is nearly the same transition temperature as the sample (without $O_2$ annealing) exposed to the air for 3 days. For comparison, we have also annealed the as-grown sample at $70°C$ in vacuum for half day, and found no significant change in R(T) data. Thus it is the oxygen that plays an important role in improving the superconductivity of the FeSeTe films. Since the film that underwent an oxygen annealing process may create some iron oxide, a disorder of iron vacancies might result and hence the $T_C$ would be enhanced [24].

The stoichiometry of the thin film was analyzed by using the Rutherford Backscattering Spectrometry (RBS), which showed the composition of the film was actually $FeSe_{0.46}Te_{0.54}$ (for simplicity, we designate this as FeSeTe in this paper). This result suggests that the Se/Te ratio in thin film is slightly higher than that of the target

($FeSe_{0.3}Te_{0.7}$) [25]. The X-ray diffractometer (XRD) data for the oxygen annealed sample is shown in Fig. 2. As can be seen, only (0 0 l) peaks are observed indicating that this film exhibits c-axis preferred orientation. Furthermore, the location of the (0 0 1) diffraction peak (~ 14.9°) can be used to infer the composition of our sample [25], which is in excellent agreement with that indicated by RBS data. Upon closer examination of the (0 0 1) diffraction peak reveals a shoulder-like structure with its center locates at slightly lower angle than the peak position. We believe that it is due to the lattice mismatch between FeSeTe and MgO substrate, the in-plane crystalline axis of the first few layers of FeSeTe above the substrate are stretched and as a consequence, the c-axis is shrunken. As the film grows thick, the dislocation lines set in and the sample returns to its usual lattice constants. Indeed, we have found that the $T_C$ of FeSeTe films thinner than 40 nm varies a lot. Besides, from our previous measurement in the patterned FeSeTe film with width about 10 μm and thickness of 40 nm, the film critical current density was around $6 \times 10^5 A/cm^2$ at 4.2 K.

### III. FABRICATION OF FESETE / AU JUNCTIONS

Using the capability in fabricating high quality epitaxial FeSeTe thin film on MgO substrate by PLD system, we fabricate FeSeTe / Au c-axis junctions to study their electric transport properties. Fig. 3(a) and 3(b) show the top view and the cross-sectional view of our junction respectively. We first in-situ deposited the FeSeTe and Au layer on MgO substrate with thickness 80 nm and 10 nm respectively. The purpose of depositing the first Au layer is to protect the FeSeTe film from oxidation so that we can have a clean SN interface as possible. Next, after standard photo-lithography technique was applied to define the first FeSeTe / Au pattern, an Argon ion-milling system was used to form the sample geometry labeled "FeSeTe" in Fig. 3(b). In order to ensure the tunneling direction in our junction is along c-axis, we used the lift off process to cover an insulating layer, 10 nm Silicon nitride (SiN), on the edge of the FeSeTe / Au stripe subsequently. We then used the thermal evaporator to deposit the 100 nm Au layer on this sample, and finally, the FeSeTe / Au tunneling junction was obtained by using another photo-lithography and Argon ion-milling process. Our junction area is about $5 \mu m \times 5 \mu m$.

### IV. EXPERIMENTAL RESULTS AND DISSCUSSION

Using a four-probe configuration, with current through leads 1 and 3, and voltage through leads 2 and 4, as shown in Fig. 3(a), we have measured the temperature dependence of the junction resistance, shown as the blue line in Fig. 4. What is shown as the red line in the same figure is the temperature dependence of the FeSeTe film measured by using a two-probe configuration, with both current and voltage leads through 1 and 2 in Fig. 3(a). From the R(T) of FeSeTe film, a clear superconducting transition was observed with the onset (90%$R_n$), offset (10%$R_n$) transition temperature and the transition width (90%$R_n$-10%$R_n$) about 13.5 K, 11.3 K, and 2.2 K respectively. On the other hand, the junction resistance rises as temperature reduces from 300K to $T_C$, then it drops precipitously at $T_C$. The insert reveals that junction resistance continuously decreases from $T_C$ to about 5 K, and then it rises again as the temperature further decreases to below 5 K.

Fig. 5 shows the conductance spectra of our c-axis junction at 2 K. The overall conductance spectrum shows a prominent zero bias conductance peak (ZBCP) with a small splitting. The ZBCP has always been observed in the conductance spectra of several batches of junctions, so it is a robust feature independent of detail junction parameters such as the junction conductance at room temperature and surface roughness. Besides, there are two conductance minima at around $\pm 4\,\text{mV}$ and $\pm 20\,\text{mV}$. Since the conductance spectrum is fairly symmetrical with respect to the bias voltage polarity, spectra for the mostly positive polarity will be presented in the subsequent figures. As shown in Fig. 6, the temperature dependence of the conductance spectrum exhibits several interesting features. First, the ZBCP becomes smaller as the temperature approaches to $T_C$ and finally disappears when the temperature is above $T_C$. Secondly, as shown in Fig. 6(b), the separation of the clear double peaks structure at 2 K gradually reduces as temperature increases, and the double peaks merges into one when the temperature increases to above 4 K. Thirdly, there are two conductance minima at around $\pm 4\,\text{mV}$ and $\pm 20\,\text{mV}$. Fig. 6(c) and (d) show the two minima in an expanded scale, and clearly these two structures move to lower voltages as temperature increases. As temperature rises to near $T_C$, both minima gradually smear out and disappear, while the normal state junction conductance shows a V-shape as depicted in the insert of Fig. 6(a). The fact that the V-shape conductance background rises for temperature higher than 11 K is probably due to FeSeTe film becoming partially normal near the junction area. It is also worth noting that there is an inflection point located at around $\pm 6\,\text{mV}$ at 2 K, which is an important clue for the determination of the superconducting gaps as discussed later.

In order to understand the behavior of the ZBCP, the ZBCP splitting, and the other features in the conductance spectrum, we have measured their magnetic field dependence at 2 K, which is shown in Fig. 7. The applied field is perpendicular to the junction surface, i.e. along the c-axis of FeSeTe. Fig. 7(a) shows that the double peaks move closer to each other and finally merge into a single peak as the magnetic field increases. To extract the position of the double peaks more precisely, we use the Fourier filtering method to remove the high frequency noise in the conductance spectra, and then take the derivative on the noise-removed spectra to determine the peak voltage values. The inset in Fig. 7(a) shows that there is a linear relationship between the voltage differences of double peaks and the applied magnetic field. In addition, the conductance minima move to lower bias voltage with increasing magnetic field. Not shown here, we have also measured the magnetic field dependence of the conductance spectra at T = 8 K and 13.5 K, and the data reveals that magnitude of ZBCP reduces and conductance minima moves to lower bias with increasing magnetic field, similar to their dependence with increasing temperature. So it is clear that those features are related to the superconducting order parameter.

Theoretically, the Andreev conductance spectra in normal metal-superconductor junctions can be calculated by using the BTK theory [26]. Using $S_\pm$ – wave symmetry of the superconducting gaps expected for the iron-based superconductors, A. A. Golubov et. al. [27] worked out an extended BTK theory to study the Andreev conductance of the N/$S_\pm$ junction. In this theory, there are two important dimensionless parameters z and $\alpha$. The former defines the barrier strength between the normal metal and superconductor, just like the same z-parameter used in the original BTK theory, while the latter is given by:

$$\alpha = \alpha_0 \frac{\phi_q(0)}{\phi_p(0)},$$

where $\alpha_0$ defines the ratio of probability amplitudes for an electron in the normal metal side tunneling into the first or second band in the superconducting side, and $\phi(0)$ is the Bloch function in two-band metal and $p$ and $q$ denotes the Fermi vector for the two bands. As indicated by A. A. Golubov et al, a bound interface state can exist when $\alpha$ is in some specific range, i.e.

$$0 \leq \alpha^2 \leq \frac{\Delta_1}{\Delta_2} \equiv \alpha_{max}^2 \text{ and } \Delta_1 \leq \Delta_2,$$

where $\alpha_{max}$ is the maximum value of $\alpha$ for the bound state to exist.
The energy level of the bound state relative to the Fermi energy depends on $\alpha$, and it is at the Fermi energy when $a = \sqrt{\Delta_1/\Delta_2}$. And if $\alpha$ is slightly smaller than $\alpha_{max}$, then the single peak at the zero voltage splits into two peaks at finite positive and negative bias voltage of the same magnitude within $\Delta_1$. (see Fig. 2 of reference [27]).

In order to compare with our experimental data, following Golubov et. al. [27], we have calculated the conductance spectrum with $\alpha$ equal to $\alpha_{max}$, as shown in Fig. 8. The prominent ZBCP is obvious, but the presence of a single peak is very sensitive to the value of $\alpha$. Double peak structure appears for $\alpha = 0.995\alpha_{max}$. The calculated conductance curve clearly shows that the two superconducting gaps correspond to the edge of the two rounded steps in the inset of Fig. 8. Thus we can identify the two rising edges, at 4 mV and 6 mV shown in Fig. 6(c), as the two superconducting gaps within the context of $S_{\pm}$-wave pair symmetry.

To have a better fitting to our experimental conductance spectrum at 2 K, we have refined our calculations by using $z = 1.8$, $\alpha = 0.8151$, $\Delta_1 = 4\,\text{meV}$, and $\Delta_2 = 6\,\text{meV}$. The results are shown in Fig. 9. We see that the overall conductance spectrum can be reproduced quite well, including the double peaks, the two rounded steps corresponding to the two superconducting gaps. Furthermore, since our temperature dependence of conductance spectra always show a clear ZBCP features, $\alpha$ value must keep almost the same ratio to form the ZBCP. Thus it implies that the temperature dependence of the two gaps must have the same functional form.

However, the experimental conductance plateau beyond the larger gap is not as high as the calculated one, and the observed minimum conductance at 20 mV is simply not present in the theory. Thus we speculate that there might be a third large gap at around 20 mV, and somehow the presence of this large gap suppresses the conductance and producing a minimum. To check this idea, we plot the voltage of this conductance minimum versus temperature shown in Fig. 10. Interestingly, the measured data points follow the BCS-gap versus temperature curve quite nicely. It is difficult to believe it is indeed a large superconducting gap. However, it is worthy to note that a large gap structure in an ultra thin FeSe film of one unit-cell thick has already been reported by Q. Y. Wang et al. by scanning tunneling microscopy (STM) [28]. Recently our own STM measurements [29] in the twisted FeSeTe grain-boundary region have also produced gaps of similar magnitude. Traditionally, the superconducting gap value is determined by the location of the conductance peaks. In the typical STM experiments, the superconducting gap is observed as a conductance peak, but in our N/S junction measurement, the gap-like features were determined by a conductance dip. Although we do not have a theory to identify the conductance minimum with a superconducting gap, it is potentially possible if we draw analogy to the extended BTK theory. Experimentally we like to point out a crucial difference between our conductance spectra and other STM experiments, namely our z factor is much smaller than those in the typical STM experiments.

## V. CONCLUSION

We have studied the temperature and magnetic field dependence of the conductance spectrum in our FeSeTe/Au c-axis junctions. A prominent ZBCP as well as some gap-like features were observed when the temperature is below the $T_C$ of FeSeTe. By using the extended nodeless $S_\pm$ − wave BTK theory with $z = 1.8$, $\alpha = 0.8151$, $\Delta_1 = 4 \text{ meV}$, and $\Delta_2 = 6 \text{ meV}$, most features of the conductance spectrum can be qualitatively explained. We argue that the temperature dependence of these two gaps must follow the same functional form. There is a discrepancy between our conductance spectra and the simulation result in the voltage range larger than $\Delta_2$ may be due to the existence of a large third gap (~20 meV). An addition, we have discovered an interesting $T_C$ enhancement effect do to oxygen-annealing effect for our pulse-laser-deposited FeSeTe thin films. The $T_C$ of the as-grown thin film can be increased up to 3 K after an oxygen-annealing process. We believe that the induced the disordered iron vacancies because of the oxidation process would result in a $T_C$ enhancement.

## ACKNOWLEDGEMENTS

This work was supported by National Science Council, Taiwan R. O. C. Grant Number: NSC 101-2112-M-007-013.

**Figure Captions**

Fig. 1. Temperature dependence of resistance for the (a) as-grown exposed to air for different days and (b) oxygen annealed FeSeTe thin film. The improvement of superconductivity may result from the oxygen in the air. The insets show the detail of the superconducting transition of these FeSeTe thin films.

Fig.. 2. The $\theta-2\theta$ X-ray diffraction pattern of the FeSeTe film. Only (0 0 l) peaks are observed indicating that this film exhibits c-axis preferred orientation. The inset shows the (0 0 1) peak ($\sim 14.9°$) of FeSeTe thin film.

Fig. 3. (a) the top view and (b) the cross-sectional view of the FeSeTe / Au c-axis junction.

Fig. 4. Temperature dependence of resistance for the FeSeTe / Au c-axis junction (shown as blue line) and the two probe FeSeTe (shown as red line). The inset shows the detail of the superconducting transition. The resistance of the two probe FeSeTe shows onset and offset $T_C$ at 13.5 K and 11 K, and the residue resistance ($\sim 1\Omega$) below $T_C$ is due to the contact resistance.

Fig. 5. The overall conductance spectra of our FeSeTe / Au c-axis junction at 2 K.

Fig. 6. (a) The overall conductance spectra of our FeSeTe / Au c-axis tunnel junction with temperature ranging from 2 K to 10.0 K and the insert shows the temperature from 10.0 K to 14.5K. We divide the conductance spectra into three regions: (b) the ZBCP region (-1~ 1 mV), (c) the low bias region (2~ 9 mV), and (d) the high bias region (10~ 30 mV). The spectrum always shows the ZBCP below $T_C$. In addition to the ZBCP, a clear double peak can be seen below 3 K. Besides, there are some gap-like features in the spectrum, which decrease with increasing temperature before they become indistinguishable in the background.

Fig. 7. The conductance spectrum of the junction with magnetic field ranging from 0 T to 9 T at 2 K. (a) shows the detail double peaks feature and the insert shows the field dependence of the location of the peaks, which follows a linear relationship. (b) shows the detail evolution of the dip structure.

Fig. 8. Conductance spectrum based on the $S_{\pm}$ model. A ZBCP without double peak splitting can be established as $\alpha$ is equal to $\alpha_{\max}$. Other parameters were kept the same

as Fig. 2 in reference [27].

Fig. 9. (a)- (c) is the conductance spectrum at 2 K in the absence of magnetic field, and (d)- (f) is the corresponding simulation curve based on the $S_\pm$ – wave model. The fitting parameter was given as $z = 1.8$, $\alpha = 0.8151$, $\Delta_1 = 4\,\text{meV}$, and $\Delta_2 = 6\,\text{meV}$. The conductance spectrum can be established qualitatively by this model.

Fig. 10. Temperature dependency of the location of the dip structures in the high bias region. The dips follow a BCS-like trend, which may indicate the third larger gap structure in the FeSeTe superconductors.

Fig. 1.

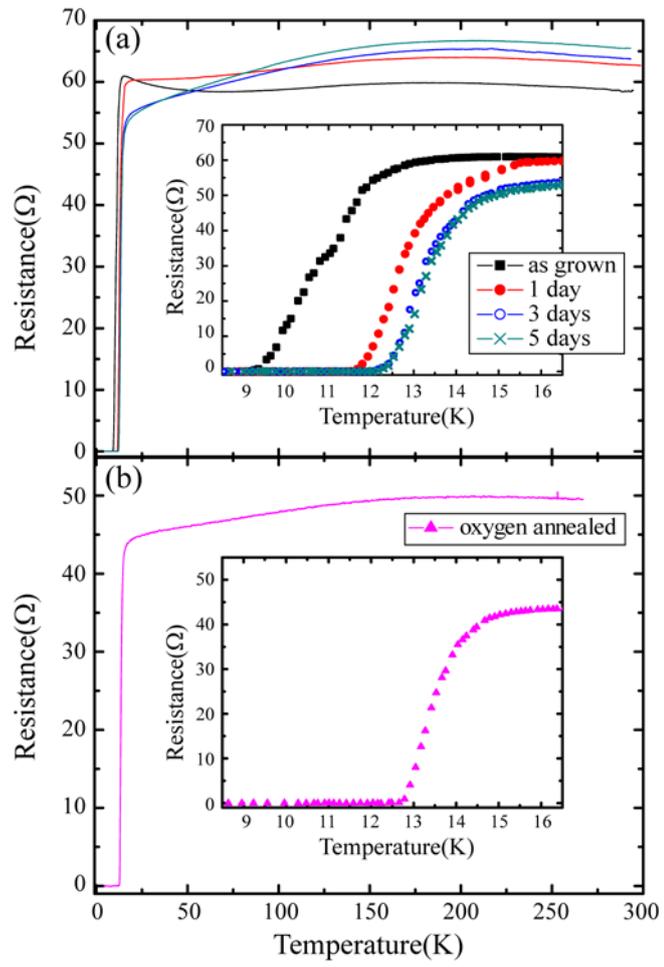

Fig. 2.

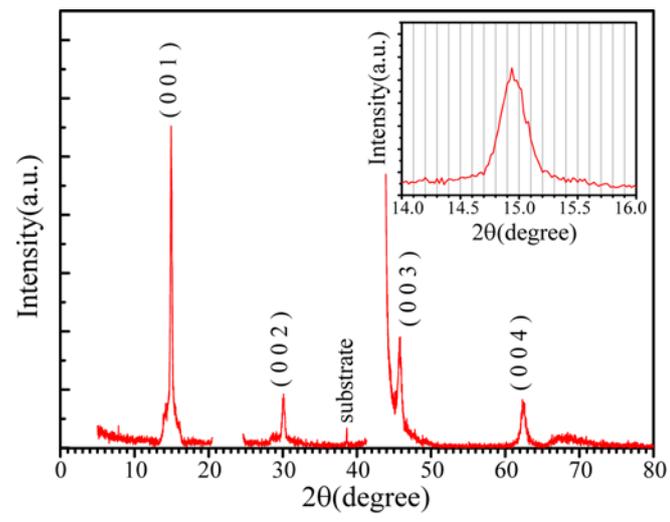

Fig. 3.

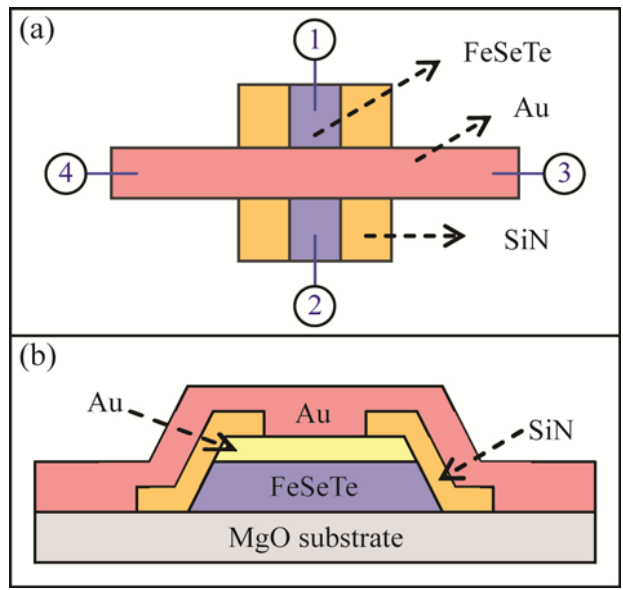

Fig. 4.

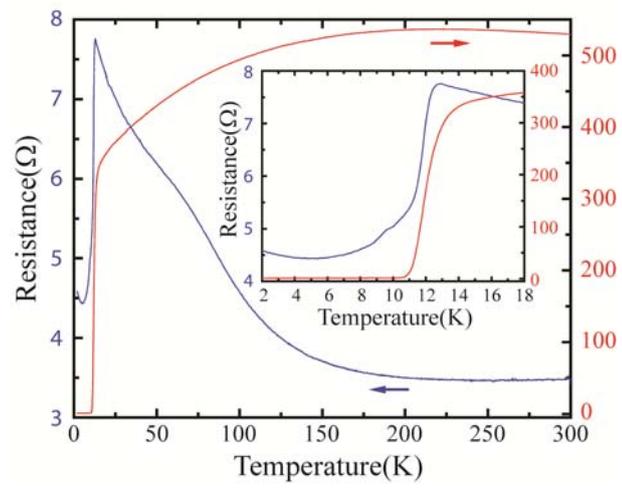

Fig. 5.

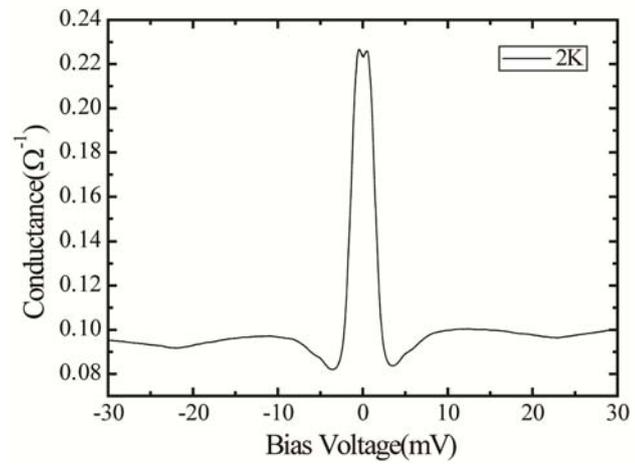

Fig. 6.

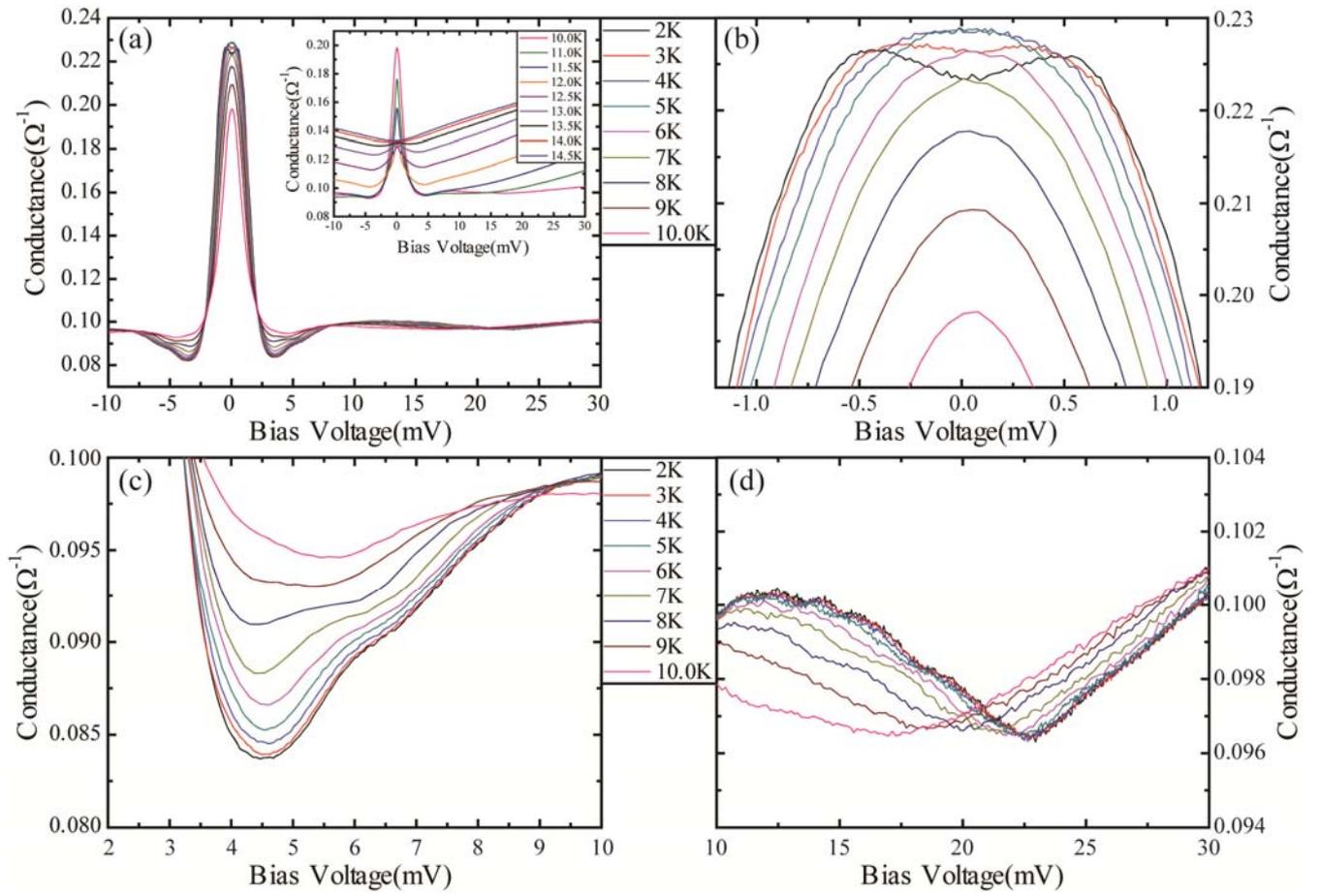

Fig. 7.

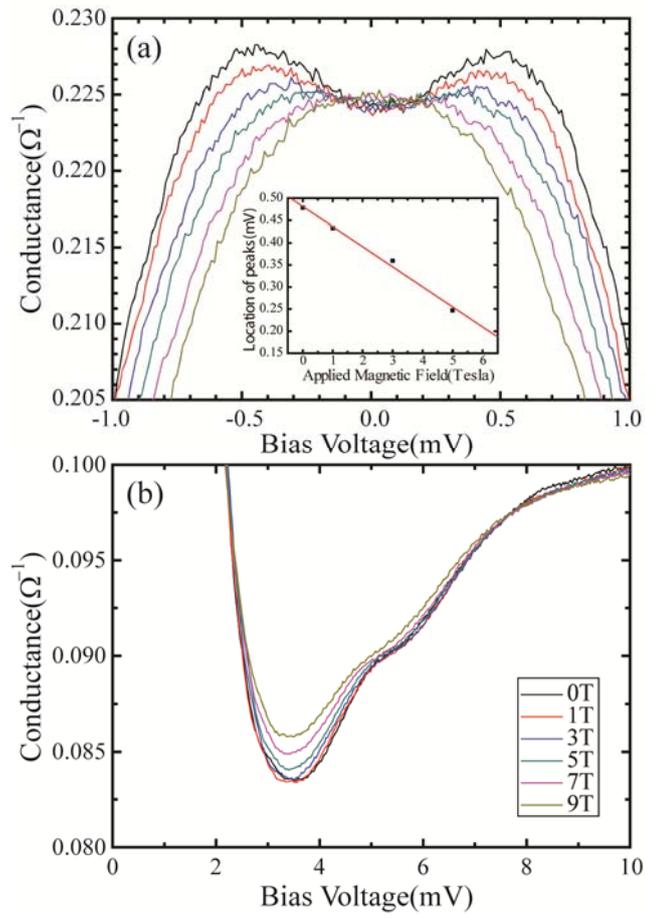

Fig. 8.

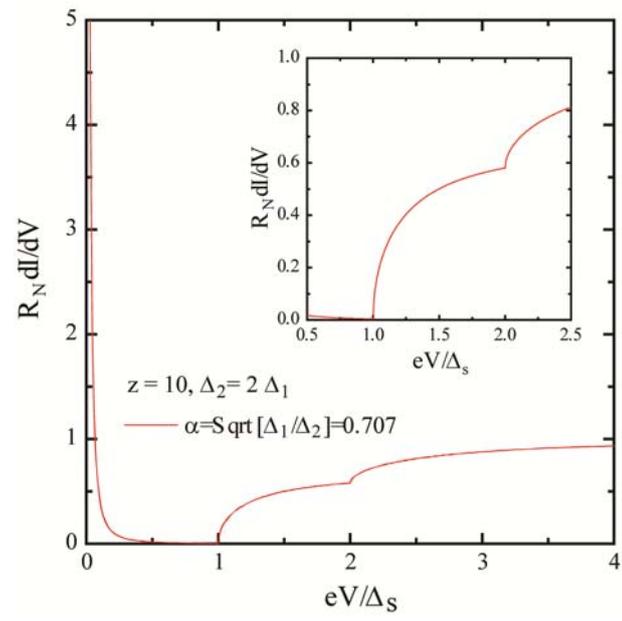

Fig. 9.

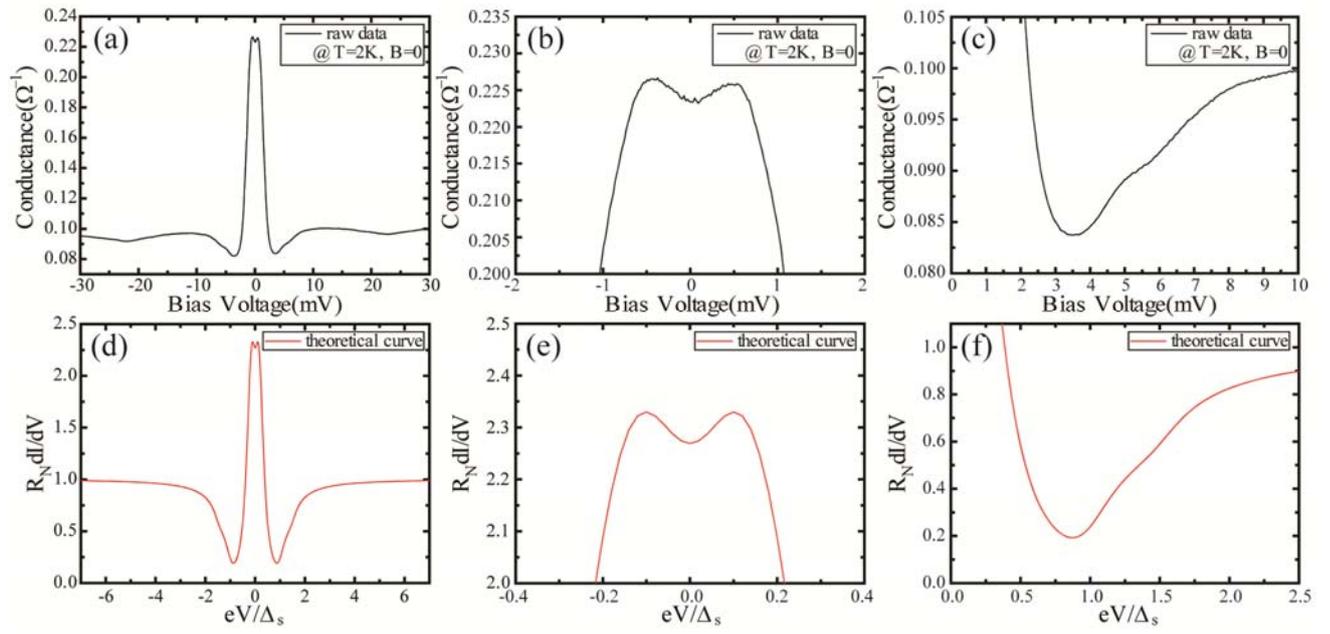

Fig. 10.

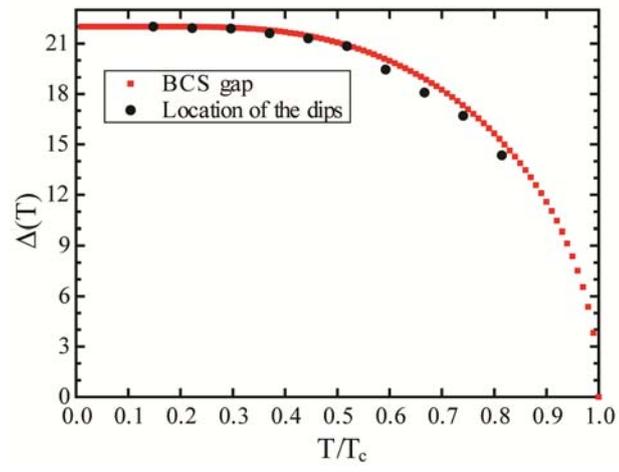